\documentclass[final]{elsart3p}

\usepackage{graphicx}
\usepackage{textcomp}
\usepackage{amssymb}
\usepackage{amsmath}

\begin{document}
\journal{J. Solid State Chem.}
\date{3 December 2007}

\begin{frontmatter}

\author{I.~M.~Ranieri} and
\ead{iranieri@ipen.br}
\author{S.~L.~Baldochi}
\ead{baldochi@ipen.br}
\address{Center for Lasers and Applications, Inst. Pesquisas Energeticas \& Nucl., \\CP 11049, Butant\~{a} 05422-970, S\~{a}o Paulo, SP, Brazil}
\author{D.~Klimm\corauthref{cor1}}
\ead{klimm@ikz-berlin.de}
\corauth[cor1]{Corresponding author: Tel.: +49 30 6392 3024; fax: +49 30 6392 3003.}
\address{Institute for Crystal Growth, Max-Born-Str. 2, 12489 Berlin, Germany}

\title{The Phase Diagram GdF$_3$--LuF$_3$}

\begin{abstract}
The phase diagram gadolinium fluoride -- lutetium fluoride was determined by differential scanning calorimetry (DSC) and X-ray powder diffraction analysis. Both pure components undergo a reversible first order transformation to a high temperature phase. The mutual solubility of both components is unlimited in the orthorhombic room temperature phase. The maximum solubility of Lu in the high temperature phase of GdF$_3$ (tysonite type) is about 20\% and the maximum solubility of Gd in LuF$_3$ ($\alpha$-YF$_3$ type) is about 40\%. Intermediate compositions of the low temperature phase decompose upon heating in a peritectoid reaction to a mixture of both high temperature phases.
\end{abstract}

\end{frontmatter}


\section{Introduction}

Rare earth trifluorides (REF$_3$, RE = La, Ce, \ldots Lu), including yttrium fluoride, were extensively studied in the past, instead of this many questions about the high-temperature polymorphic transformations and its structure so far remain. Oftedal \cite{Oftedal29,Oftedal31} and Schlyter \cite{Schlyter53}, determined the LaF$_3$ structure as hexagonal with the $P\,6_3/mcm$ space group and two formula units per elementary cell, using synthesized powder and the lanthanum mineral tysonite (= fluocerite). Mansmann \cite{Mansmann64,Mansmann65} and at same time Zalkin et al. \cite{Zalkin66} established the current accepted structure of the LaF$_3$. Using smaller X-ray wavelength (Mo$K_\alpha$) it was possible to observe weaker additional reflections on a single crystal, and the structure of the LaF$_3$ tysonite structure was described as trigonal with space group $P\,\bar{3}c1$ and $Z=6$. The light rare earths fluorides from La to Nd crystallize in this structure. All other rare earth fluorides crystallize at room temperature in the orthorhombic structure determined by Zalkin and Templeton \cite{Zalkin53} for YF$_3$, also referred as $\beta$-YF$_3$, space group $P\,nma$ and $Z=4$.

Another characteristic pointed out by Mansmann for the change of the structure from trigonal to orthorhombic was due to a relation between the size of the metal and its coordination number. When the ionic radius of the rare earth $r_\mathrm{RE}$ decreases, the fluorine ions tend to touch each other, resulting in a repulsive energy. A critical ratio $r_\mathrm{RE}/r_\mathrm{F}=0.94$ was proposed for the change from the LaF$_3$ to the $\alpha$-YF$_3$ structure, taking into account the ionic radius calculated by Ahrens \cite{Ahrens52}. Furthermore, the trigonal structure could be stabilized if the REF$_3$-orthorhombic structure becomes deficient in fluorine ions or at high temperatures \cite{Mansmann65}.

Thoma et al. \cite{Thoma66} studied the behavior of the REF$_3$ from SmF$_3$ to LuF$_3$ taking into account X-ray diffraction at high temperature and differential thermal analysis (DTA) experiments. It was observed that the REF$_3$ from Sm to Ho have the trigonal $P\,\bar{3}c1$ high-$T$ structure of LaF$_3$ ``tysonite'' type mentioned above. For the small rare earths from Er to Lu upon heating, the $\beta$-YF$_3$ structure changes to a hexagonal not well identified $\alpha$-YF$_3$ structure. This structure was considered tentatively isostructural with the trigonal $\alpha$-UO$_3$ structure by Sobolev et al. \cite{Sobolev73,Sobolev76c}, belonging to $P\,\bar{3}m1$ ($D^3_{3d}$ ) space group, $Z=1$ \cite{Zachariasen48}. Nevertheless, the $\alpha$-UO$_3$ structure is not yet well understood too. Either superstructures or a lower (e.g. orthorhombic) symmetry were discussed, as single crystal show a biaxial optical interference figure \cite{Greaves72,Siegel71}.

Jones and Shand \cite{Jones68} proved that it was possible to grow crystals of the four tysonites, LaF$_3$, CeF$_3$, PrF$_3$ and NdF$_3$, but only the orthorhombic DyF$_3$ and HoF$_3$ using CdF$_2$ as scavenger. After Garton \cite{Garton78} and Pastor \cite{Pastor74e} it was established that GdF$_3$, TbF$_3$, DyF$_3$ and HoF$_3$ crystals can be grown from oxygen free compounds and in a reactive atmosphere, confirming the thermodynamic studies by Spedding et al \cite{Spedding74}.

For the intermediate SmF$_3$, EuF$_3$, and GdF$_3$ two subsequent transformations $P\,nma \stackrel{(1)}{\longleftrightarrow} P\,\bar{3}c1 \stackrel{(2)}\longleftrightarrow P\,6_3/mmc$ were discussed by Greis \cite{Greis85} (the number in brackets indicates the order of the phase transformation [PT]). These PT were inferred from electron diffraction experiments with LaF$_3$, where small synthesized crystals presented also sub-cells reflections as observed by Schlyter \cite{Schlyter53} and Maximov \cite{Maximov85}. Stankus \cite{Stankus00} claimed that at high $T$ the $\beta$-YF$_3$ type ($P\,nma$) and the LaF$_3$ type (tysonite, $P\,\bar{3}c1$) become practically identical. Sobolev et al. \cite{Sobolev77} constructed the phase diagrams of the systems GdF$_3$--LnF$_3$ (Ln = Tb, Ho, Er, Yb), solid solutions regions were proposed without phase transitions in all systems, when the cation mean ionic radius was between that of the Tb$^{3+}$ and Er$^{3+}$.

Recently, the present authors have published a phase diagram study of the system GdF$_3$--YF$_3$ \cite{Klimm07b}. It was established that both components undergo a solid-state phase transformation of first order before melting. Additionally, a $\lambda$-shaped maximum of $c_\mathrm{p}(T)$ being characteristic for a second order transformation was found for GdF$_3$. Both low-$T$ and high-$T$ phases exhibit unlimited mutual solubility. This observation raises the question, whether the high-$T$ structures of YF$_3$ (reported as $P\,\bar{3}m1$) and GdF$_3$ (reported as $P\,6_3/mmc$) may really be different \cite{Sobolev73}.

In the current paper, the phase diagram of the system GdF$_3$--LuF$_3$ is reported for the first time. The interest in this phase diagram derived from DTA studies regarding the phase diagram of the system LiF--Gd$_{1-x}$Lu$_x$F$_3$, which is interesting to develop new solid-solution crystals of the type LiGd$_{1-x}$Lu$_x$F$_4$ to be used as laser host.


\section{Experimental}

Mixtures of Gd$_{1-x}$Lu$_x$F$_3$ with $x = 0.2, 0.4, 0.6$ and $0.8$, respectively, were prepared using commercial LuF$_3$ (AC Materials, 6N purity) and GdF$_3$ synthesized from commercial Gd$_2$O$_3$ powder (Alfa, 5N purity) by hydrofluorination. The oxide was placed in a platinum boat inside a platinum tube, and slowly heated in a stream of argon gas (White Martins, purity 99.995\%) and HF gas (Matheson Products, purity 99.99\%) up to $850^{\,\circ}$C. This process is described in detail elsewhere \cite{Guggenheim63,Ranieri01}. Conversion rates $>99.9$\% of the theoretical value calculated for the reaction Gd$_2$O$_3$ + 6\,HF $\longrightarrow$ 2\,GdF$_3$ + 3\,H$_2$O were measured by comparing the masses prior to and after the hydrofluorination process. These samples were used to the DSC measurements.

Thermoanalytic measurements were performed with a NETZSCH STA 409CD with DSC/TG sample carrier (thermocouples type S). The sample carrier was calibrated for $T$ and sensitivity at the phase transformation points of BaCO$_3$ and at the melting points of Zn and Au. Sample powders ($50-70$~mg) were placed in graphite DSC crucibles with lid. As graphite is not wetted by the molten fluorides, the melt forms one single almost spherical drop (diameter $d\approx3$~mm) that could be used for the subsequent X-ray phase analysis.

The vapor pressures of both fluorides at their melting points are high. Fortunately, the evaporation rate for pure GdF$_3$ or LuF$_3$, respectively, was found to be almost identical. Thus it can be assumed that the partial evaporation does not lead to a considerable concentration shift. The inhomogeneous powder samples were homogenized in a first heating/cooling cycle with $\pm$40\;K/min. Here the heating was performed to that $T_\mathrm{max}$ where the DSC melting peak was just finished and the molten sample could homogenize. Depending on $x$, this was the case for $1145^{\:\circ}$C$\leq T_\mathrm{max} \leq1280^{\:\circ}$C and due to the large heating rate the mass loss did never exceed 4\% in these preliminary mixing cycles. Without intermediate opening of the apparatus, the mixing cycle was followed by a measuring run with a heating rate of 10\,K/min. Although the crucicles were covered by lids, the evaporation of $\approx10-15$\% sample mass during this DSC/TG run cannot be avoided. Cooling curves showed often supercooling and were not used for the construction of the phase diagram. In total, 14 different compositions ranging from pure GdF$_3$ ($x=0$) to pure LuF$_3$ ($x=1$) were measurered.

Other samples were melted under a flux of hydrogen fluoride gas, then pulverized to be analyzed by powder X-ray diffraction, using a Bruker AXS diffractometer, model D8 Advance, operated at 40\,kV and 30\,mA, in the $2\,\theta$ range of $22.5-68.5^\circ$. The diffraction patterns where treated with the Rietveld Method \cite{Rietveld69} using the GSAS program to calculate the lattice parameters \cite{Larson07}.


\section{Results and Discussion}

It turned out that GdF$_3$ as well as LuF$_3$ showed similar DSC heating curves: A first endothermal peak due to the first order PT ($T_\mathrm{PT}^{\mathrm{GdF}_3}=902^{\,\circ}$C or $T_\mathrm{PT}^{\mathrm{LuF}_3}=946^{\,\circ}$C, respectively) is followed by a second endothermal peak due to melting ($T_\mathrm{f}^{\mathrm{GdF}_3}=1252^{\,\circ}$C or $T_\mathrm{f}^{\mathrm{LuF}_3}=1182^{\,\circ}$C, respectively). The values for GdF$_3$ were measured and compared with literature data recently \cite{Klimm07b}, and for lutetium fluoride one finds values $943\leq T_\mathrm{PT}^{\mathrm{LuF}_3}\leq963$ and $1180\leq T_\mathrm{f}^{\mathrm{LuF}_3}\leq1188$ ($T$ in $^\circ$C) in the literature \cite{Thoma66,Greis85,Stankus00}. Only Jones \& Sand \cite{Jones68} reported the very low value $T_\mathrm{PT}^{\mathrm{LuF}_3}\approx874^{\,\circ}$C. Like in the recent study \cite{Klimm07b} weak $\lambda$ shaped bends were found in the DSC curves of GdF$_3$ rich mixtures with $x<0.2$. This could be an indication for a second order transformation between $T_\mathrm{PT}$ and $T_\mathrm{f}$ but will not be discussed here.

\begin{figure}[htb]
\centering
\includegraphics[width=0.45\textwidth]{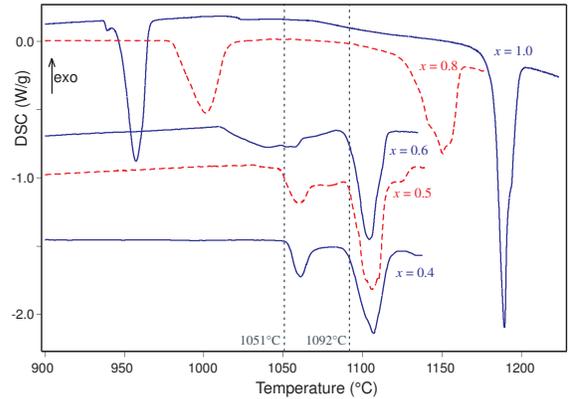}
\caption{DSC curves (2nd heating with 10\,K/min) of LuF$_3$ ($x=1.0$) and of four mixtures Gd$_{1-x}$Lu$_x$F$_3$ with $0.4\leq x\leq0.8$. $T_\mathrm{PT}=1051^{\,\circ}$C and $T_\mathrm{eut}=1092^{\,\circ}$C are the two isothermal phase boundaries in Fig.~\ref{fig:GdF3-LuF3}.}
\label{fig:DSC}
\end{figure}

Fig.~\ref{fig:DSC} shows DSC heating curves for five compositions starting from pure LuF$_3$ ($x=1.0$) to Gd$_{0.6}$Lu$_{0.4}$F$_3$. The PT peak for $x=1.0$ is disturbed by a shoulder on the low $T$ side that is assumed to result from some contamination of the sample, as no additional peak that could be related to the presence of oxyfluorides was detected on cooling. With increasing GdF$_3$ content the melting peak shifts to lower $T$ until it reaches for $x\approx0.6$ the constant value $T_\mathrm{eut}=1092^{\,\circ}$C. On the contrary, the PT peak shifts with increasing GdF$_3$ content to higher $T$ and reaches for $x\approx0.5$ the constant value $T_\mathrm{PT}=1051^{\,\circ}$C. The same $T_\mathrm{PT}$ and $T_\mathrm{eut}$ are reached if one starts with pure GdF$_3$ and adds LuF$_3$.

\begin{figure}[htb]
\centering
\includegraphics[width=0.45\textwidth]{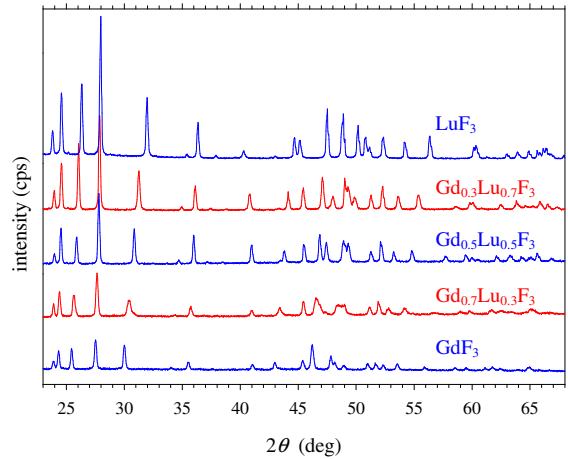}
\caption{Diffractions patterns of some Gd$_{1-x}$Lu$_x$F$_3$ samples utilized to determine the lattice parameters at room temperature.}
\label{fig:X-ray}
\end{figure}

\begin{figure}[htb]
\centering
\includegraphics[width=0.40\textwidth]{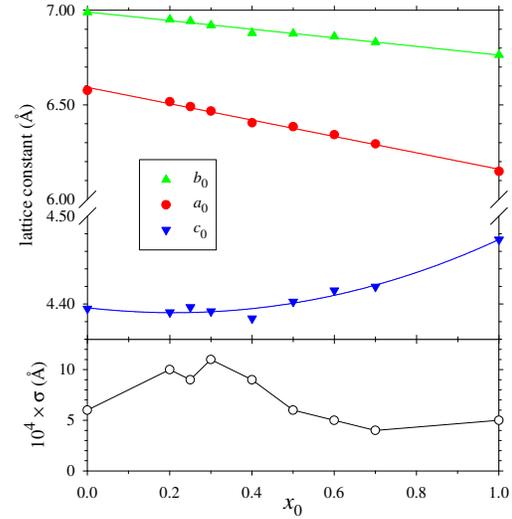}
\caption{Top: lattice constants for the $P\,nma$ phase ($\beta$-YF$_3$ type) of Gd$_{1-x}$Lu$_x$F$_3$ solid solutions at room temperature. Bottom: experimental error $\sigma = \mathit\Delta a_0 + \mathit\Delta b_0 + \mathit\Delta c_0$.}
\label{fig:lattice_par}
\end{figure}

The lattice constants of nine samples with initial LuF$_3$ concentrations $0\leq x_0\leq 1$ were measured by X-ray diffraction. All diffraction patterns were fitted considering the orthorhombic structure with the $P\,nma$ space group, no parasitic peaks that would indicate the presence of other phases were observed at room temperature (Fig.~\ref{fig:X-ray}). The experimental errors $\lbrace \mathit\Delta a_0, \mathit\Delta b_0, \mathit\Delta c_0\rbrace$ were always $\leq5\times10^{-4}$\,\AA\ but were found to be maximum in the region $0.2\leq x_0\leq 0.4$. Fig.~\ref{fig:lattice_par} shows the total error $\sigma = \mathit\Delta a_0 + \mathit\Delta b_0 + \mathit\Delta c_0$ versus the LuF$_3$ concentration $x_0$ of the sample, together with $a_0, b_0, c_0$.

Table~\ref{tab:X-ray} reports the results for calculated values of $a_0$, $b_0$ and $c_0$, the volume of the unit cell $V$, the mass density $\varrho$ and the mean rare-earth ionic radius in the solid solution $\bar{r}=(1-x_0)r^{[8]}_\mathrm{Gd}+x_0 \,r^{[8]}_\mathrm{Lu}$. $r^{[8]}_\mathrm{Gd}$ and $r^{[8]}_\mathrm{Lu}$ are the ionic radii of Gd$^{3+}$ and Lu$^{3+}$, respectively, with coordination number 8 as determinated by Shannon et al. \cite{Shannon76}.

\begin{table}[htb]
\caption{Lattice parameters for the $P\,nma$ phase of Gd$_{1-x}$Lu$_x$F$_3$ $(0\leq x\leq1)$}
\begin{tabular}{lllllll}
\hline
\tiny Lu content & $\bar{r}$    & $a_0$              &    $b_0$           & $c_0$              & $V$                 & $\varrho$ \normalsize          \\
\tiny($x$, molar)    &        (\AA) & (\AA)              &    (\AA)           & (\AA)              & (\AA$^3$)           &  (g/cm$^3$) \normalsize      \\
\hline
\tiny
0.00       &              & 6.571$^\mathrm{a}$ & 6.984$^\mathrm{a}$ & 4.390$^\mathrm{a}$ & 201.47$^\mathrm{a}$ & 7.063$^\mathrm{a}$ \\
0.00       & 1.053        & 6.5758(2)          & 6.9898(2)          & 4.3947(2)          & 201.999(11)         & 7.045 \\
0.20       & 1.038        & 6.5143(6)          & 6.9517(5)          & 4.3907(3)          & 198.833(31)         & 7.157 \\
0.25       & 1.034        & 6.4908(4)          & 6.9429(3)          & 4.3964(2)          & 198.125(18)         & 7.331 \\
0.30       & 1.030        & 6.4671(5)          & 6.9209(4)          & 4.3918(2)          & 196.569(28)         & 7.359 \\
0.40       & 1.023        & 6.4102(3)          & 6.8867(2)          & 4.3885(1)          & 193.733(13)         & 7.588 \\
0.50       & 1.015        & 6.3838(2)          & 6.8770(2)          & 4.4025(1)          & 193.275(8)          & 7.667 \\
0.60       & 1.007        & 6.3421(2)          & 6.8615(2)          & 4.4154(2)          & 192.143(11)         & 7.773 \\
0.70       & 1.000        & 6.2855(2)          & 6.8222(1)          & 4.4140(1)          & 189.282(10)         & 7.967 \\
1.00       & 0.977        & 6.1481(2)          & 6.7640(2)          & 4.4732(1)          & 186.021(11)         & 8.283 \\
1.00       &              & 6.150$^\mathrm{b}$ & 6.760$^\mathrm{b}$ & 4.468$^\mathrm{b}$ & 185.83$^\mathrm{b}$ & 8.291$^\mathrm{b}$ \normalsize\\
\hline
\end{tabular} \\
$^\mathrm{a}$ PDF 49-1804; $^\mathrm{b}$ PDF 32-0612
\label{tab:X-ray}
\end{table}

The lattice parameters $a_0$ and $b_0$ drop linearly with $x_0$ and $c_0$ is almost constant for $x_0\leq0.4$ and starts then to rise slightly up to the value of pure LuF$_3$ (Fig.~\ref{fig:lattice_par}). The smooth variation of $a_0$, $b_0$ and $c_0$ over the whole concentration range is obviously the result of unlimited mutual solubility of GdF$_3$ and LuF$_3$ in their low-$T$ phases. The variation of $c_0$ in this system is very similar to the $\vec{c}$ axis variation of the pure rare earth fluorides. In the orthorhombic rare-earth fluorides due to the lanthanides contraction, there is a linear decreasing in the $c_0$ parameter value from 4.40\,\AA\ (SmF$_3$) down to 4.376\,\AA\ (DyF$_3$) and then a nearly exponential increase up to 4.467\,\AA\ (LuF$_3$) \cite{Zalkin53}. Taking into account the ionic radius of Ho$^{3+}$, namely 1.015\,\AA\ (with a coordination number of 8) \cite{Shannon76}, and the mean ionic radius of the composition Gd$_{0.5}$Lu$_{0.5}$F$_3$, one has the same value.

\begin{figure}[htb]
\centering
\includegraphics[width=0.45\textwidth]{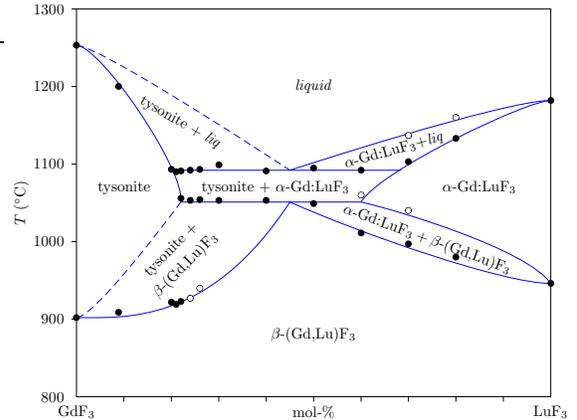}
\caption{Phase diagram GdF$_3$--LuF$_3$. \textbullet\ -- sharp peaks. \textopenbullet\ -- small peaks or from offsets. $\alpha, \beta$ mean high-$T$ phase or low-$T$ phase, respectively.}
\label{fig:GdF3-LuF3}
\end{figure}

The DSC and XRD results are summarized in Fig.~\ref{fig:GdF3-LuF3} where full circles represent experimental DSC points that could be well determined by extrapolated onsets of sharp peaks. Such points represent the lower boundary of PT regions or, in the case of melting, the solidus line. The higher boundary of PT regions or the liquidus line, respectively, are less remarkable as here the PT process or melting just finishes and the DSC curve returns to the basis line. Such more vague determined events are marked by hollow circles in Fig.~\ref{fig:GdF3-LuF3}.

Two horizontal lines at $1051\pm2^{\,\circ}$C and $1092\pm2^{\,\circ}$C are the most prominent feature of the experimental phase diagram Fig.~\ref{fig:GdF3-LuF3}. Obviously, two different processes with strong thermal effects start independent on composition $x$ for a broad range $0.2\lesssim x\lesssim0.5\ldots0.6$ always at the same $T_\mathrm{PT}=1051^{\,\circ}$C or $T_\mathrm{eut}=1092^{\,\circ}$C, respectively. The components GdF$_3$ and LuF$_3$ show complete miscibility in the liquid phase. If the melt is cooled, either Gd-saturated LuF$_3$ ($\alpha$-YF$_3$ type, $P\,\bar{3}m1$) or Lu-saturated GdF$_3$ (tysonite type, $P\,6_3/mmc$ or $P\,\bar{3}c1$) crystallize from the melt. Both high-$T$ phases belong to different space groups and unlimited miscibility is therefore prohibited --- instead a eutectic is formed below $1092^{\,\circ}$C. In the eutectic mixture between $T_\mathrm{eut}$ and $T_\mathrm{PT}$ both constituents are mutually saturated. In the $\beta$-phase both components exhibit unlimited mutual solubility. The phase transformation occurring at $T_\mathrm{PT}$ for medium concentrations $x$
\begin{equation}
\mathrm{tysonite} + \alpha\mathrm{-YF}_3 \rightleftarrows \beta\mathrm{-YF}_3
\label{eq:peri}
\end{equation}
is a peritectoid reaction.

Near the left or right rims of the phase diagram, single phase Lu-doped GdF$_3$ (tysonite structure) as well as single phase Gd-doped LuF$_3$ (= $\alpha$-Gd:LuF$_3$) transform upon cooling to $\beta$-(Gd,Lu)F$_3$, crossing a 2-phase field where $\alpha$- and $\beta$-phase are mixed and the upper and lower limit of this 2-phase field depend on $x$. Both 2-phase fields ``tysonite + liq.'' and ``tysonite + $\beta$-(Gd,Lu)F$_3$'' are broader than the corresponding fields on the LuF$_3$ rich side where tysonite is replaced by $\alpha$-Gd:LuF$_3$. The larger widths of the 2-phase fields on the GdF$_3$ rich side may result in stronger phase segregation of such compositions, resulting after cooling to room temperature in microscopic grains with a wider variation of compositions. As different compositions of the Gd$_{1-x}$Lu$_x$F$_3$ grains result in different lattice constants, the larger experimental error $\sigma$ that is observed for Gd-rich mixtures (but not for pure GdF$_3$) can be explained by variations of $x$ resulting from segregation.


\section{Conclusion}

Solid solutions Gd$_{1-x}$Lu$_x$F$_3$ are formed over the whole concentration range $0\leq x\leq1$ and crystallize in the same $\beta$-YF$_3$ type structure with space group $P\,nma$ like the pure components GdF$_3$ and LuF$_3$. Upon heating GdF$_3$ undergoes a first order phase transformation to a tysonite type structure whereas LuF$_3$ undergoes a first order transformation to a $\alpha$-YF$_3$ type structure that is not isomorphous to tysonite. As a result, a miscibility gap separates the high-$T$ phases of GdF$_3$ and LuF$_3$. For intermediate compositions $0.2\lesssim x\lesssim0.6$ the phase transformation from single phase Gd$_{1-x}$Lu$_x$F$_3$ to a mixture of the high-$T$ phases of GdF$_3$ (saturated with Lu) and LuF$_3$ (saturated with Gd) is a peritectoid decomposition.

The behavior of the system GdF$_3$--LuF$_3$ is somewhat uncommon: Usually the mutual solubility of components rises with $T$, but in the present case both components show unlimited miscibility in the low-$T$ phase only. It should be noted that Nafziger et al. \cite{Nafziger73} obtained similar results (constant eutectic and phase transformation temperatures $T_\mathrm{eut}=1073^{\,\circ}$C and $T_\mathrm{PT}=1045^{\,\circ}$C, respectively, over an extended composition range $\gtrsim50$\%) with DTA measurements in the LaF$_3$--YF$_3$ system. Nafziger et al. speculated that this is an indication for a low-$T$ intermediate phase, but the present authors think that a peritectoid decomposition (\ref{eq:peri}) as shown in Fig.~\ref{fig:GdF3-LuF3} of this work would be a more simple explanation and that the postulation of an (otherwise not proven) intermediate phase is not necessary.


\ack{

The authors acknowledge financial support from CAPES (grant number 246/06) and DAAD (grant number D/05/30364) in the framework of the PROBRAL program.}


\end{document}